\documentclass[conference]{IEEEtran}
\IEEEoverridecommandlockouts
\usepackage[numbers]{natbib}
\usepackage{algorithmic}
\usepackage{graphicx}
\usepackage{flushend}
\usepackage{dblfloatfix}
\usepackage{textcomp}
\usepackage{xcolor}
\usepackage{url}
\usepackage{booktabs}
\usepackage{tcolorbox}
\usepackage{multirow}
\usepackage{ulem}
\usepackage{balance}
\usepackage{colortbl}
\usepackage{listings}
\usepackage{enumitem}
\usepackage{comment}
\usepackage{colortbl}
\usepackage{amssymb}
\usepackage{rotating}
\usepackage{longtable}
\usepackage{tabularx,booktabs}
\usepackage{lipsum}

\newlist{UP}{enumerate}{1}
\setlist[UP]{label=U\textsubscript{\arabic*}:}
\newlist{RP}{enumerate}{1}
\setlist[RP]{label=R\textsubscript{\arabic*}:}
\newlist{CP}{enumerate}{1}
\setlist[CP]{label=C\textsubscript{\arabic*}:}
\newlist{FP}{enumerate}{1}
\setlist[FP]{label=F\textsubscript{\arabic*}:}

\usepackage[percent]{overpic}%
\usepackage{tikz}
\usepackage{moresize}
\usepackage{caption}
\usepackage{subcaption}
\colorlet{punct}{red!60!black}
\definecolor{background}{HTML}{EEEEEE}
\definecolor{delim}{RGB}{20,105,176}
\colorlet{numb}{magenta!60!black}

\lstdefinelanguage{json}{
	basicstyle=\scriptsize,
	numberstyle=\scriptsize,
	stepnumber=1,
	numbersep=5pt,
	showstringspaces=false,
	breaklines=true,
	frame=lines,
	backgroundcolor=\color{background},
	literate=
	*{0}{{{\color{numb}0}}}{1}
	{1}{{{\color{numb}1}}}{1}
	{2}{{{\color{numb}2}}}{1}
	{3}{{{\color{numb}3}}}{1}
	{4}{{{\color{numb}4}}}{1}
	{5}{{{\color{numb}5}}}{1}
	{6}{{{\color{numb}6}}}{1}
	{7}{{{\color{numb}7}}}{1}
	{8}{{{\color{numb}8}}}{1}
	{9}{{{\color{numb}9}}}{1}
	{:}{{{\color{punct}{:}}}}{1}
	{,}{{{\color{punct}{,}}}}{1}
	{\{}{{{\color{delim}{\{}}}}{1}
	{\}}{{{\color{delim}{\}}}}}{1}
	{[}{{{\color{delim}{[}}}}{1}
	{]}{{{\color{delim}{]}}}}{1},
}

\definecolor{maroon}{cmyk}{0, 0.87, 0.68, 0.32}
\definecolor{halfgray}{gray}{0.55}
\definecolor{ipython_frame}{RGB}{207, 207, 207}
\definecolor{ipython_bg}{RGB}{247, 247, 247}
\definecolor{ipython_red}{RGB}{186, 33, 33}
\definecolor{ipython_green}{RGB}{0, 128, 0}
\definecolor{ipython_cyan}{RGB}{64, 128, 128}
\definecolor{ipython_purple}{RGB}{170, 34, 255}

\lstdefinelanguage{python}{
	morekeywords={access,and,break,class,continue,def,del,elif,else,except,exec,finally,for,from,global,if,import,in,is,lambda,not,or,pass,print,raise,return,try,while},
	morekeywords=[2]{abs,all,any,basestring,bin,bool,bytearray,callable,chr,classmethod,cmp,compile,complex,delattr,dict,dir,divmod,enumerate,eval,execfile,file,filter,float,format,frozenset,getattr,globals,hasattr,hash,help,hex,id,input,int,isinstance,issubclass,iter,len,list,locals,long,map,max,memoryview,min,next,object,oct,open,ord,pow,property,range,raw_input,reduce,reload,repr,reversed,round,set,setattr,slice,sorted,staticmethod,str,sum,super,tuple,type,unichr,unicode,vars,xrange,zip,apply,buffer,coerce,intern},
	sensitive=true,
	morecomment=[l]\#,
	morestring=[b]',
	morestring=[b]",
	morestring=[s]{'''}{'''},
	morestring=[s]{"""}{"""},
	morestring=[s]{r'}{'},
	morestring=[s]{r"}{"},
	morestring=[s]{r'''}{'''},
	morestring=[s]{r"""}{"""},
	morestring=[s]{u'}{'},
	morestring=[s]{u"}{"},
	morestring=[s]{u'''}{'''},
	morestring=[s]{u"""}{"""},
	literate=
	{á}{{\'a}}1 {é}{{\'e}}1 {í}{{\'i}}1 {ó}{{\'o}}1 {ú}{{\'u}}1
	{Á}{{\'A}}1 {É}{{\'E}}1 {Í}{{\'I}}1 {Ó}{{\'O}}1 {Ú}{{\'U}}1
	{à}{{\`a}}1 {è}{{\`e}}1 {ì}{{\`i}}1 {ò}{{\`o}}1 {ù}{{\`u}}1
	{À}{{\`A}}1 {È}{{\'E}}1 {Ì}{{\`I}}1 {Ò}{{\`O}}1 {Ù}{{\`U}}1
	{ä}{{\"a}}1 {ë}{{\"e}}1 {ï}{{\"i}}1 {ö}{{\"o}}1 {ü}{{\"u}}1
	{Ä}{{\"A}}1 {Ë}{{\"E}}1 {Ï}{{\"I}}1 {Ö}{{\"O}}1 {Ü}{{\"U}}1
	{â}{{\^a}}1 {ê}{{\^e}}1 {î}{{\^i}}1 {ô}{{\^o}}1 {û}{{\^u}}1
	{Â}{{\^A}}1 {Ê}{{\^E}}1 {Î}{{\^I}}1 {Ô}{{\^O}}1 {Û}{{\^U}}1
	{œ}{{\oe}}1 {Œ}{{\OE}}1 {æ}{{\ae}}1 {Æ}{{\AE}}1 {ß}{{\ss}}1
	{ç}{{\c c}}1 {Ç}{{\c C}}1 {ø}{{\o}}1 {å}{{\r a}}1 {Å}{{\r A}}1
	{€}{{\EUR}}1 {£}{{\pounds}}1
	{^}{{{\color{ipython_purple}\^{}}}}1
	{=}{{{\color{ipython_purple}=}}}1
	{+}{{{\color{ipython_purple}+}}}1
	{*}{{{\color{ipython_purple}$^\ast$}}}1
	{/}{{{\color{ipython_purple}/}}}1
	{+=}{{{+=}}}1
	{-=}{{{-=}}}1
	{*=}{{{$^\ast$=}}}1
	{/=}{{{/=}}}1,
	literate=
	*{-}{{{\color{ipython_purple}-}}}1
	{?}{{{\color{ipython_purple}?}}}1,
	identifierstyle=\color{black}\ttfamily,
	commentstyle=\color{ipython_cyan}\ttfamily,
	stringstyle=\color{ipython_red}\ttfamily,
	keepspaces=true,
	showspaces=false,
	showstringspaces=false,
	rulecolor=\color{ipython_frame},
	numberstyle=\tiny\color{halfgray},
	backgroundcolor=\color{ipython_bg},
	basicstyle=\scriptsize,
	keywordstyle=\color{ipython_green}\ttfamily,
}

\def\BibTeX{{\rm B\kern-.05em{\sc i\kern-.025em b}\kern-.08em
	T\kern-.1667em\lower.7ex\hbox{E}\kern-.125emX}}

\newcommand\RqOne{$RQ_1$}
\newcommand\RqTwo{$RQ_2$}
\newcommand\RqOneSen{(\RqOne) Are generated codes similar to human codes?}
\newcommand\RqTwoSen{(\RqTwo) Can generated codes perform better than human codes?}

\begin{document}

\title{An Empirical Evaluation of Competitive Programming AI: A Case Study of AlphaCode}

\author{\IEEEauthorblockN{Sila Lertbanjongngam$^1$, Bodin Chinthanet$^2$, Takashi Ishio$^2$, Raula Gaikovina Kula$^2$, \\ Pattara Leelaprute$^1$, Bundit Manaskasemsak$^1$, Arnon Rungsawang$^1$, Kenichi Matsumoto$^2$}
	\IEEEauthorblockA{$^1$\textit{Department of Computer Engineering, Faculty of Engineering, Kasetsart University, Bangkok, Thailand}\\
		$^2$\textit{Graduate School of Science and Technology, Nara Institute of Science and Technology, Nara, Japan}\\
		\{sila.l, bundit.m, arnon.r\}@ku.th, \{bodin.ch, ishio, raula-k, matumoto\}@is.naist.jp, pattara.l@ku.ac.th}
}

\maketitle

\begin{abstract}
AlphaCode is a code generation system for assisting software developers in solving competitive programming problems using natural language problem descriptions.
Despite the advantages of the code generating system, the open source community expressed concerns about practicality and data licensing.
However, there is no research investigating generated codes in terms of  code clone and performance.
In this paper, we conduct an empirical study to find code similarities and performance differences between AlphaCode-generated codes and human codes.
The results show that (i) the generated codes from AlphaCode are similar to human codes (i.e., the average maximum similarity score is 0.56) and
(ii) the generated code performs on par with or worse than the human code in terms of execution time and memory usage.
Moreover, AlphaCode tends to generate more similar codes to humans for low-difficulty problems (i.e., four cases have the exact same codes).
It also employs excessive nested loops and unnecessary variable declarations for high-difficulty problems, which cause low performance regarding our manual investigation.
The replication package is available at \url{https:/doi.org/10.5281/zenodo.6820681}
\end{abstract}

\begin{IEEEkeywords}
	code generation, code similarity, code performance
\end{IEEEkeywords}

\section{Introduction}
\label{sec:introduction}
Over the past few years, Artificial Intelligence (AI) applications have become very popular, especially in the software engineering field, as they help software developers to work faster and more efficiently~\citep{chakkrit:ASR2021}.
The examples of tasks that AI can assist developers in the software development process are (i) software defect prediction \citep{Tantithamthavorn:ICSE2016, Okutan:EMSE2014}, (ii) cost estimation \citep{Samson:IST1997, Huang:IST2015}, (iii) task prioritization \citep{Perini:TSE2012}, (iv) expert recommendation \citep{Xuan:ICSE2012}, (v) security vulnerability detection \citep{Zheng:ICSE-SEIP2021, Nguyen:ASE2021}, and (vi) code generation \citep{chen:2021, Li:2022}.
These AI applications rely on massive amounts of data from software development tools such as version control systems, issue tracking systems, and continuous integration and deployment systems (CI/CD).

In the case of the code generation system, AI models are expected to synthesize new and unseen codes, while they know existing programs from their training dataset~\citep{allamanis_adverse_2019}.
Recently, several works focus on evaluating code generation systems on Codex \citep{chen:2021} and GitHub Copilot \citep{Web:GitHubCopilot} from security \citep{Pearce:SP2022} and correctness aspects \citep{Nguyen:MSR2022}.
These code generation systems require a code comment or a short natural language as input for solving the given task.

AlphaCode, a transformer-based AI code generation system developed by DeepMind \citep{Li:2022}, is currently the state of the art for competitive programming AI.
The advantage of AlphaCode compared to other code generation systems is that it can generate source codes from competitive programming problem descriptions, which usually require an understanding of algorithms and complex natural languages.
However, as the model behind the AI used human codes for the training process, concerns about data-related issues were raised in the open source community, such as data privacy, licensing, and data extraction attacks \citep{Carlini:USENIX2021, Web:GitHubCopilotRecitation}.
As there is also no research investigating AlphaCode, it is unclear whether generated codes are clones of human codes.
It is also unclear whether the performance of the generated codes at which, if deployed, how many resources (i.e., time and memory) are required to run the program.

In this paper, we conduct an empirical study to find code similarities and performance differences between AlphaCode-generated codes and human codes.
We define two following research questions:
\begin{itemize}
	\item \RqOneSen
	\item \RqTwoSen
\end{itemize}
We collected 44 samples of valid generated codes in C++ and Python languages from the AlphaCode official website~\citep{Web:alphacode} that solve 22 problems on Codeforces.
We then retrieved 31,736 human codes by using the Codeforces API.
For \RqOne, we compare source code similarity between the generated and human codes using metrics to detect source code reuse~\citep{hata_same:2021}.
For \RqTwo, we measure the performance of the codes based on an existing performance comparison work~\citep{Leelaprute:ICPC2022}.
		
The results show that (i) the generated codes from AlphaCode are similar to the human codes (i.e., the average maximum similarity score is 0.56 and 0.50 for C++ and Python), while the code fragments in the generated code are comprised of various human codes (i.e., the uniqueness of 3.30\% and 8.94\% for C++ and Python) and
(ii) the generated code performs on par with or worse than the human code in terms of execution time and memory usage.
Furthermore, AlphaCode generates code that is more similar to human code for low-difficulty problems (i.e., four cases with the same code) and employs excessive nested loops and unnecessary variable declarations (i.e., using \textit{long long} instead of \textit{int}, unused integer list), resulting in poor performance.

The rest of this paper is organized as follows.
Section~\ref{sec:background} describes the background of this research.
Section~\ref{sec:setup} shows our analysis method.
Section~\ref{sec:results} describes results of the analysis.
Section~\ref{sec:threats} discusses limitations and threats to validity.
Section~\ref{sec:conclusion} concludes this paper and suggests future work.

\section{Background}
\label{sec:background}
In this section, we briefly explain terms that have been used throughout the paper.

\subsection{Competitive Programming}

\begin{figure}
	\centering
	\includegraphics[width=.45\textwidth]{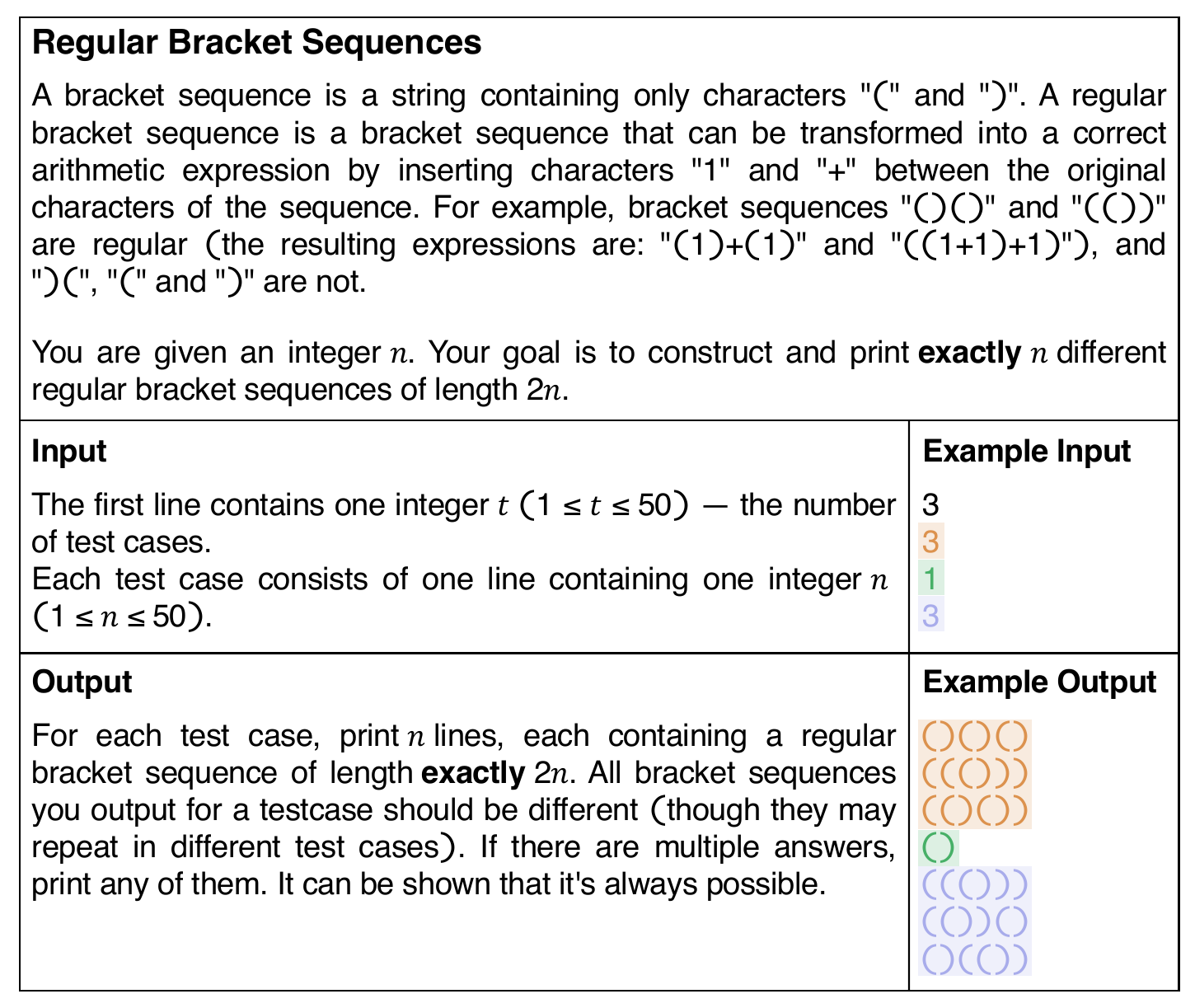}
	\caption{Example competitive programming problem from Codeforces (1574A - Regular Bracket Sequences) \citep{Web:CodeforcesExample}.}
	\label{fig:example-problem}
	\vspace{-1em}
\end{figure}
    Competitive programming is a competition that gives well-known computer science problems to contestants and asks them to solve those problems by writing source codes as quickly as possible \citep{Halim:2013}.
    To solve a problem, contestants have to analyze the description and use their logical, mathematical, and computer science skills to find the optimal solution, which is usually not directly stated in the description text.
    There are several well-known programming competitions that are annually held and supported by software organizations, such as Google Code Jam \citep{Web:GLCodeJam}, Meta Hacker Cup \citep{Web:FBHackerCup}, and ACM ICPC \citep{Web:ACMICPC}.
    The format of competitions can be different depending on the host; for example, the number of problems, competition times, and available programming languages.
    In this paper, we focus on Codeforces, one of the most popular competitive programming platforms that has over one million users and regularly holds a weekly competition \citep{Web:CodeforcesBlog}.
    
    Figure \ref{fig:example-problem} shows the competitive programming problem example, called \textit{Regular Bracket Sequences}, hosted on Codeforces.
    The problem description consists of (i) the details of the problem, and (ii) input and output constraints, which include the format and the example test case. 
    In this example, the contestants are asked to generate \textit{n} different bracket sequences that can be transformed into a correct expression for each test case.
    There are multiple solutions to this problem, but they have to be executed within limited memory and time.

\subsection{AlphaCode: A Competitive-level AI Code Generation}
    In February 2022, DeepMind introduced AlphaCode, a large-scale transformer-based code generation system.
    The main target of AlphaCode is to generate code for solving competitive programming problems that require an understanding of algorithms and complex natural languages.
    One of the highlights of AlphaCode is that it can generate the entire program from a long natural language description compared to Codex, a code generation system used in GitHub Copilot \citep{chen:2021, Web:GitHubCopilot}, which is capable of generating code for a simple task with a short solution (e.g., function, API usage).
    
    Several studies focused on the evaluation of the transformer-based code generation system.
    \citet{chen:2021} evaluated Codex and found that it has a strong performance for easy interview problems.
    \citet{Nguyen:MSR2022} evaluated the suggested codes of GitHub Copilot with LeetCode and found that it achieved at most 57\% of correctness score.
    \citet{Pearce:SP2022} found that generated code can introduce the security vulnerability issue.
    In the case of AlphaCode, the evaluation is still limited as only DeepMind provides such information \citep{Li:2022}.
    They found that AlphaCode achieved an average ranking within the top 54.3\% with over 5,000 human contestants.
    Additionally, AlphaCode achieved 20.36\% and 7.75\% solve rates (i.e., pass@k metric) in introductory and competition tasks.
    However, the understanding of generated codes from AlphaCode is still unclear especially in terms of the quality of codes compared to the actual human codes. 
    Hence, this study empirically evaluates the code similarities and performance differences between generated codes and human codes, as described in the following section.

\section{Experiment Setup}
\label{sec:setup}
This section presents the dataset and analysis methods to understand how generated codes are similar to human codes. 

\begin{table}[]
\centering
\caption{Summary information of our dataset.}
\label{tab:dataset}
\begin{tabular}{l|ccc|c} 
\toprule
\multicolumn{5}{c}{\textbf{Summary information of collected generated codes}}                                   \\ 
\midrule
\multicolumn{1}{c|}{\textbf{Detail}}          & \textbf{Mean} & \textbf{Median} & \textbf{SD} & \textbf{Total}  \\ 
\hline
\multicolumn{1}{c|}{\textbf{C++ language}}    &               &                 &             &                 \\
- \# codes                           & 1             & 1               & 1           & 22              \\
- \# line of code                              & 31.772        & 26              & 11.688      & 699             \\
\multicolumn{1}{c|}{\textbf{Python language}} &               &                 &             &                 \\
- \# codes                           & 1             & 1               & 1           & 22              \\
- \# lines of code                                & 16.318        & 15.5            & 6.724       & 359             \\ 
\midrule
\multicolumn{5}{c}{\textbf{Summary information of collected human codes}}                                       \\ 
\midrule
\multicolumn{1}{c|}{\textbf{Detail}}          & \textbf{Mean} & \textbf{Median} & \textbf{SD} & \textbf{Total}  \\ 
\hline
\multicolumn{1}{c|}{\textbf{C++ language}}    &               &                 &             &                 \\
- \# codes                              & 977           & 1000            & 102.48      & 21,508           \\
- \# lines of code                                & 52.117        & 36.0            & 48.740      & 1,120,933         \\
\multicolumn{1}{c|}{\textbf{Python language}} &               &                 &             &                 \\
- \# codes                              & 464           & 461             & 352.19      & 10,228           \\
- \# lines of code                                & 14.274        & 11.0            & 16.990      & 145,998          \\
\bottomrule
\end{tabular}
\end{table}

\subsection{{Dataset}}
    Our dataset consists of (i) the generated codes from AlphaCode and (ii) the human codes from Codeforces.
    From the AlphaCode official website~\cite{Web:alphacode}, there are 141 generated codes from 43 different problems written in C++ and Python languages.
    However, some of these generated codes do not successfully solve the problems.
    Additionally, some problems have generated code in only one language.
    In this paper, we consider only problems that have both C++ and Python generated codes and those that can solve the problems correctly.
    We crawled and collected the 44 generated codes that solve 22 problems written in C++ and Python languages.

    To empirically evaluate the generated codes, we retrieved human codes from Codeforces using a provided API on May 17, 2022.
    However, we were able to retrieve at most 1,000 submitted codes per language for each problem due to API limitations.
    We also collected the difficulty score of each problem.
    In the end, 21,508 C++ and 10,228 Python human codes were collected for our experiment.
    The summary of our dataset is described in Table \ref{tab:dataset}.
		
\subsection{Analysis for \RqOne}

\begin{table}[h]
\centering
\caption{An example of similarity value. A bold trigram is unique to a code fragment.}
\label{tab:codexample}
\begin{tabular}{c|c}

    \toprule
    \textbf{Code fragment a}                                                                     & \textbf{Code fragment b}                                                \\ \midrule  

  if ( x != y ); x++; & 	 if ( y == x ); y++;  \\ \midrule
    
    \textbf{trigrams(a)}                                                                     & \textbf{trigrams(b)}                                                      \\
    \midrule
    \textless \_, \_, if \textgreater{},  \textless \_, if, ( \textgreater{},                & \textless \_, \_, if \textgreater{},  \textless \_, if, ( \textgreater{}, \\
    \textbf{\textless if, (, x \textgreater{}}, \textbf{\textless (, x, != \textgreater{}},  & \textless if, (, y \textgreater{},  \textless (, y, == \textgreater{},    \\
    \textbf{\textless x, !=, y \textgreater{}},  \textbf{\textless !=, y, ) \textgreater{}}, & \textless y, ==, x \textgreater{},  \textless ==, x, ) \textgreater{},    \\
    \textbf{\textless y, ), ; \textgreater{}}, \textbf{\textless ), ;, x \textgreater{}},    & \textless x, ), ; \textgreater{},  \textless ), ;, y \textgreater{},      \\
    \textbf{\textless ;, x, + \textgreater{}},  \textbf{\textless x, +, + \textgreater{}},   & \textless ;, y, + \textgreater{},  \textless y, +, + \textgreater{},      \\
    \textless +, +, ; \textgreater{},  \textless +, ;, \_ \textgreater{},                    & \textless +, +, ; \textgreater{},  \textless +, ;, \_ \textgreater{},     \\
    \textless ;, \_, \_ \textgreater{},                                                      & \textless ;, \_, \_ \textgreater{},                                       \\
\midrule
    \multicolumn{2}{l}{$sim(a, b) = \frac{5}{21} = 0.238, ~unique(a, \{b\}) = \frac{8}{13} = 0.615$} \\
    \bottomrule

\end{tabular}

\end{table}

To analyze the source code similarity between generated and human codes, we make a comparison for each problem at the file-level granularity.
We have employed two source code similarity metrics that have been used to measure the degree of source code reuse~\cite{hata_same:2021}.
\begin{eqnarray*}
sim\left( f_{1}, f_{2} \right) & =  & \frac{\left| trigrams\left( f_{1} \right) \cap trigrams\left( f_{2} \right)\right|}{\left| trigrams\left( f_{1} \right)\cup  trigrams\left( f_{2} \right) \right|} \\
unique\left( f, H \right) & = & \frac{\left| trigrams\left( f \right) \setminus \bigcup_{h \in H} trigrams\left( h \right)\right|}{\left| trigrams\left( f \right) \right|} 
\end{eqnarray*}
where $trigrams(f)$ is a multiset of token trigrams (three consecutive tokens) extracted from a file $f$.
Table~\ref{tab:codexample} shows example values of the metrics for a pair of code fragments (i.e., a sample of lines in a file).

We calculate source code similarity for each pair of generated and human codes.
A higher similarity indicates that a larger amount of source code could be reused from the human code.
As Juergens et al.~\cite{juergens_code_2010} reported that different authors write different source codes, the similarity is expected to be low.

The uniqueness of a generated code is defined as the ratio of trigrams that are unique to the generated code.  In other words, it measures the number of trigrams in the generated code that are never included in human codes.
A higher uniqueness indicates that the generated code includes only its own code.  
The uniqueness becomes low if token trigrams in generated codes are also found in human codes.

\subsection{Analysis for \RqTwo}
To analyze the performance difference between generated codes and human codes, we implement the system to measure the execution time and memory usage similar to \citet{Leelaprute:ICPC2022}.
We first generate the input data based on the constraints of Codeforces problems.
Due to the limitations of the Codeforces API, it is impossible to retrieve the largest input data for each problem.
Hence, we use the upper-bond constraint described in the problem description as the size of the input in order to simulate the worst-case scenario.
We then compile and execute both generated and human codes by using our input data.
In this case, we use only human codes that are the most similar to generated codes for each problem.
We repeatedly execute codes 100 times to reduce the threat from non-deterministic and other factors that affect the execution time and memory usage.
We also set an execution time limit of five seconds to avoid the infinity loop case.

In our experiment, we use \textit{memory-profiler}, a Python library for measuring the amount of memory consumption for each code execution with the execution timestamp \citep{Web:mem-profiler}.
For compiler and interpreter setup, we use the GNU G++20 11.2.0 compiler with \textit{-O2} option for C++ code compilation and CPython 3.9.7 for Python code execution.
Our machine uses an AMD Ryzen Threadripper 3970X CPU and 128 GB of DDR4 RAM on Ubuntu 20.04 operating system.

To statistically validate the differences between the execution time and memory usage of generated codes and human codes, we use the independent sample t-test, which compares the means between two groups \citep{Ross:2017}.
Our hypothesis is that ``the execution time and memory usage between generated and human codes are the same or not".
We also measure the effect size using Cohen's d, which shows differences based on means and standard deviations of two groups \citep{Cohen:1988}. 
The interpretation is listed as follows: (i) 0 $\leq$ d \textless~0.1 as Very small, (ii) 0.1 $\leq$ d \textless~0.35 as Small, (iii) 0.35 $\leq$ d \textless~0.65 as Medium, (iv) 0.65 $\leq$ d \textless~0.9 as Large, or (v) d $\geq$ 0.9 as Very large.
In our experiment, we use NumPy \citep{Numpy:2020}, Scipy \citep{SciPy:2020}, and researchpy \citep{Web:researchpy} for the statistical test.

\begin{figure}
	\centering
	\begin{subfigure}[]{0.5\textwidth}
		\centering
		\includegraphics[width=\textwidth]{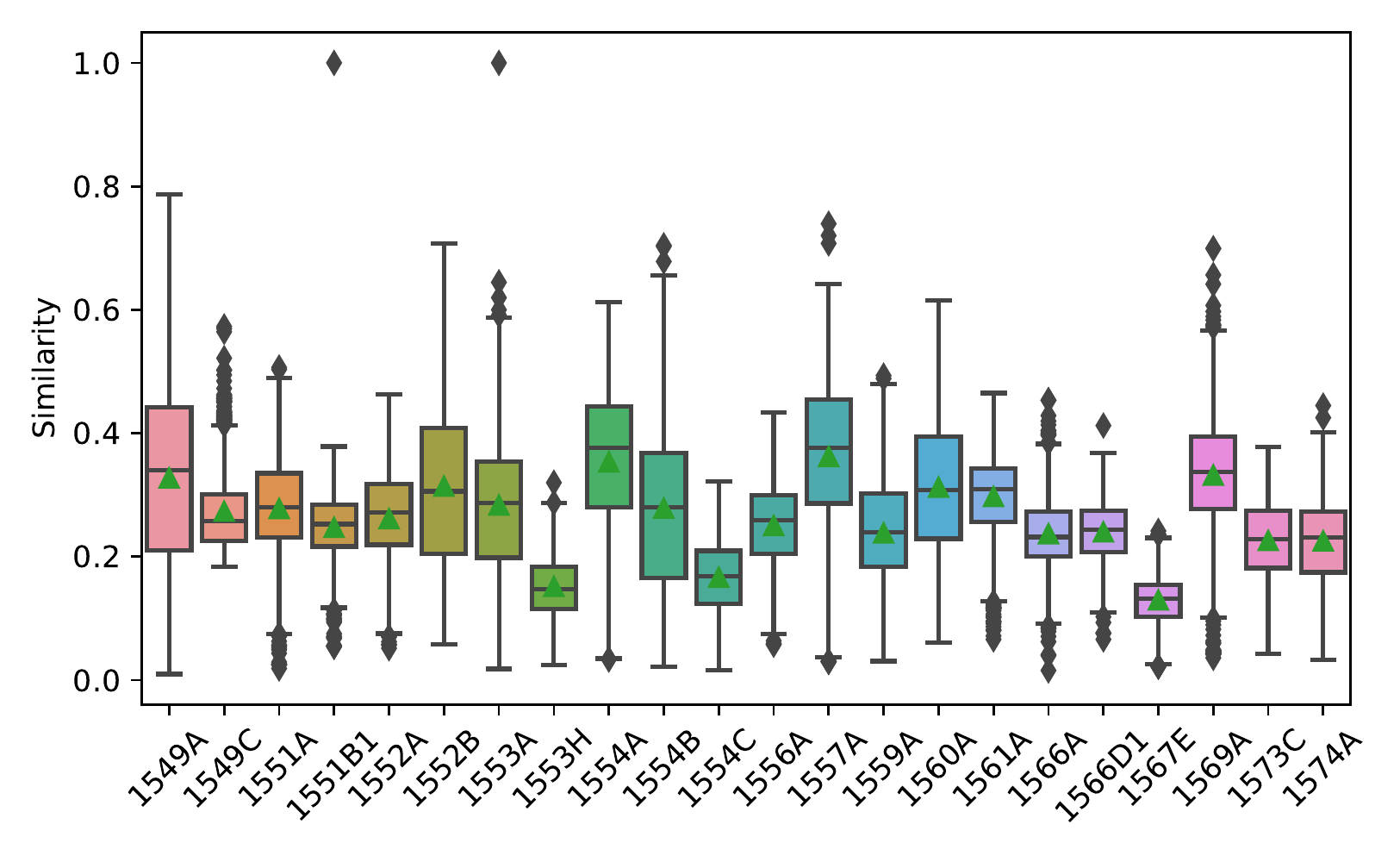}
		\vspace{-2.5em}
		\caption{C++ language}
		\label{fig:RQ1_similarity_cpp}
	\end{subfigure}
	\begin{subfigure}[]{0.5\textwidth}
		\centering
		\includegraphics[width=\textwidth]{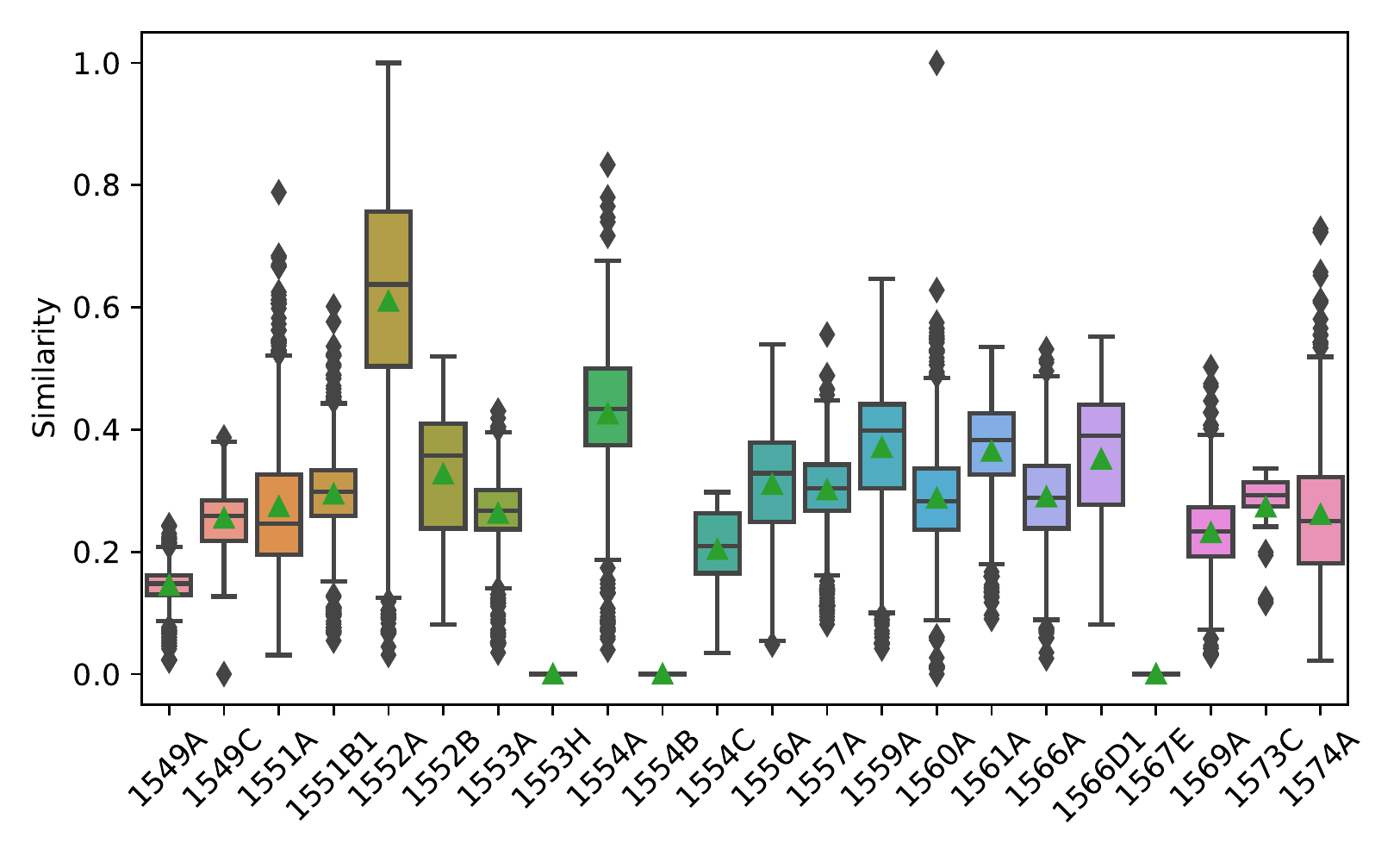}
		\vspace{-2.5em}
		\caption{Python language}
		\label{fig:RQ1_similarity_python}
	\end{subfigure}
	\caption{Similarity between generated codes and human codes.}
	\label{fig:RQ1_similarity}
\end{figure}

\section{Results}
\label{sec:results}
\subsection{\RqOneSen}

	\textbf{Similarity between generated and human codes.}
	Figure \ref{fig:RQ1_similarity} shows the similarity score between generated codes and human codes for each problem in C++ and Python.
	We find that, overall, AlphaCode can generate similar codes to human as the mean of maximum similarity scores for C++ (Figure \ref{fig:RQ1_similarity_cpp}) and Python (Figure \ref{fig:RQ1_similarity_python}) are 0.56 and 0.50 respectively.
	Only four cases that generated codes are exactly the same as human codes for both languages.
	We also find that these four cases have difficulty score at 800, which is one of the lowest score among all problems in our study.
	
\begin{table}[h]
	\centering
	\caption{Percentage of code fragment uniqueness of generated codes} 
	\label{tab:RQ1_uniqueness}
	\begin{tabular}{cccccc}
		\toprule
		\textbf{Languages} & \textbf{Mean} & \textbf{Median} & \textbf{Min} & \textbf{Max} & \textbf{SD} \\
		\midrule
		\textbf{C++}       & 3.30\%        & 2.60\% & 0\% & 11.36\%     & 3.21\%      \\
		\textbf{Python}    & 8.94\%        & 3.24\% & 0\% & 31.75\%         & 10.66\%     \\
		\bottomrule
	\end{tabular}
	\vspace{-1.4em}
\end{table}

	\textbf{Uniqueness of generated code fragments.}
	Table \ref{tab:RQ1_uniqueness} shows percentage of uniqueness of generated codes compared to human codes.
	We find that the code fragments in generated codes are common in human codes as the mean of uniqueness for C++ language is 3.30\% and 8.93\% for Python language.
    Considering these two results, it indicates that the AlphaCode model might not directly clone the training data to solve the competitive programming problems.
    However, AlphaCode generates code that is a mixture of multiple human codes.

\begin{tcolorbox}
	\textbf{Answer to \RqOne:} Yes, generated codes from AlphaCode can be similar to human codes (the average maximum similarity is 0.56). 
	The code fragments used by AlphaCode are common across all human codes.
\end{tcolorbox}

\begin{table}
	\centering
	\caption{\small Summary statistics and comparison of execution times (in second) between generated codes and human codes written in C++. Note that we show $|\Delta|$ and Cohen's d if generated codes perform significant difference (p-value $<$ 0.001).}
	\label{tab:RQ2_time_cpp}
	\scalebox{0.8}{
	\begin{tabular}{cccccccc} 
		\toprule
		\multirow{2}{*}{} & \multicolumn{1}{l}{\multirow{2}{*}{\textbf{Difficulty}}} & \multicolumn{2}{c}{\textbf{AlphaCode}}                   & \multicolumn{2}{c}{\textbf{Human}}                 & \multirow{2}{*}{$|\Delta|$} & \multirow{2}{*}{\textbf{Cohen's d}}  \\ 
		\cline{3-6}
		       & \multicolumn{1}{l}{} & \textbf{Mean}                              & \textbf{SD} & \textbf{Mean}                        & \textbf{SD} &            &              \\ 
		\midrule
		1549A  & 800                  & 0.0009                                     & 0.0022      & 0.0017                               & 0.0005      & \textbf{-} & \textbf{-}   \\
		1549C  & 1400                 & {\cellcolor[rgb]{0.753,0.753,0.753}}5.821  & 0.0069      & {\cellcolor[rgb]{1,0.8,0.404}}0.0591 & 0.0191      & 5.7619     & 399.22*  \\
		1551A  & 800                  & {\cellcolor[rgb]{0.753,0.753,0.753}}5.8216 & 0.0047      & {\cellcolor[rgb]{1,0.8,0.404}}0.0133 & 0.0027      & 5.8083     & 1501.59* \\
		1551B1 & 800                  & 0.0067                                     & 0.0018      & 0.0065                               & 0.0011      & \textbf{-} & \textbf{-}   \\
		1552A  & 800                  & 0.004                                      & 0.0006      & 0.0041                               & 0.0008      & \textbf{-} & \textbf{-}   \\
		1552B  & 1500                 & 5.8219                                     & 0.0057      & {\cellcolor[rgb]{1,0.8,0.404}}0.0463 & 0.0017      & 5.775      & 1377.4*  \\
		1553A  & 800                  & 0.0019                                     & 0.0004      & 0.0019                               & 0.0005      & \textbf{-} & \textbf{-}   \\
		1553H  & 2900                 & {\cellcolor[rgb]{0.753,0.753,0.753}}5.821  & 0.0038      & {\cellcolor[rgb]{1,0.8,0.404}}0.0899 & 0.0017      & 5.731      & 1961.69* \\
		1554A  & 800                  & {\cellcolor[rgb]{0.753,0.753,0.753}}5.8199 & 0.0048      & {\cellcolor[rgb]{1,0.8,0.404}}0.0589 & 0.0025      & 5.761      & 1487.15* \\
		1554B  & 1700                 & {\cellcolor[rgb]{0.753,0.753,0.753}}5.82   & 0.0045      & {\cellcolor[rgb]{1,0.8,0.404}}0.0203 & 0.0017      & 5.799      & 1683.69* \\
		1554C  & 1800                 & {\cellcolor[rgb]{0.753,0.753,0.753}}5.82   & 0.0052      & {\cellcolor[rgb]{1,0.8,0.404}}0.009  & 0.0017      & 5.811      & 1488.17* \\
		1556A  & 800                  & {\cellcolor[rgb]{0.408,0.796,0.816}}0.0032 & 0.0006      & 0.0104                               & 0.0016      & 0.0072     & 5.74*    \\
		1557A  & 800                  & 0.0914                                     & 0.0047      & 0.0911                               & 0.0031      & \textbf{-} & \textbf{-}   \\
		1559A  & 900                  & 0.0024                                     & 0.0027      & 0.002                                & 0.0005      & \textbf{-} & \textbf{-}   \\
		1560A  & 800                  & 0.001                                      & 0.0026      & 0.0007                               & 0.0005      & \textbf{-} & \textbf{-}   \\
		1561A  & 800                  & 0.0038                                     & 0.0066      & 0.0034                               & 0.0006      & \textbf{-} & \textbf{-}   \\
		1566A  & 800                  & {\cellcolor[rgb]{0.753,0.753,0.753}}5.8194 & 0.0041      & {\cellcolor[rgb]{1,0.8,0.404}}0.0146 & 0.0024      & 5.8048     & 1722.22* \\
		1566D1 & 1100                 & 0.0271                                     & 0.0038      & {\cellcolor[rgb]{1,0.8,0.404}}0.0121 & 0.0018      & 0.015      & 5.0*     \\
		1567E  & 2200                 & {\cellcolor[rgb]{0.753,0.753,0.753}}5.8213 & 0.0103      & {\cellcolor[rgb]{1,0.8,0.404}}1.4461 & 0.0044      & 4.3752     & 548.34*  \\
		1569A  & 800                  & {\cellcolor[rgb]{0.408,0.796,0.816}}0.0024 & 0.002       & 0.0049                               & 0.0008      & 0.0025     & 1.7*     \\
		1573C  & 1800                 & {\cellcolor[rgb]{0.408,0.796,0.816}}0.0291 & 0.0033      & 0.0327                               & 0.0036      & 0.0036     & 1.05*    \\
		1574A  & 800                  & 0.0038                                     & 0.0017      & {\cellcolor[rgb]{1,0.8,0.404}}0.0021 & 0.0004      & 0.0017     & 1.33*    \\
		\bottomrule
	\end{tabular}
	}
\end{table}

\begin{table}
	\centering
	\caption{\small Summary statistics and comparison of execution times (in second) between generated codes and human codes written in Python. Note that we show $|\Delta|$ and Cohen's d if generated codes perform significant difference (p-value $<$ 0.001).}
	\label{tab:RQ2_time_python}
	\scalebox{0.8}{
	\begin{tabular}{cccccccc} 
		\toprule
		\multirow{2}{*}{} & \multicolumn{1}{l}{\multirow{2}{*}{\textbf{Difficulty}}} & \multicolumn{2}{c}{\textbf{AlphaCode}}                   & \multicolumn{2}{c}{\textbf{Human}}                 & \multirow{2}{*}{$|\Delta|$} & \multirow{2}{*}{\textbf{Cohen's d}}  \\ 
		\cline{3-6}
		       & \multicolumn{1}{l}{} & \textbf{Mean}                              & \textbf{SD} & \textbf{Mean}                        & \textbf{SD} &            &              \\ 
		\midrule
		1549A  & 800                  & 0.2069                                     & 0.0067      & 0.2082                               & 0.0056      & \textbf{-} & \textbf{-}   \\
		1549C  & 1400                 & {\cellcolor[rgb]{0.753,0.753,0.753}}5.8217 & 0.0035      & {\cellcolor[rgb]{1,0.8,0.404}}0.6428 & 0.0084      & 5.1789     & 796.38*  \\
		1551A  & 800                  & 0.2271                                     & 0.0037      & 0.2272                               & 0.0084      & \textbf{-} & \textbf{-}   \\
		1551B1 & 800                  & 0.2208                                     & 0.0052      & 0.2203                               & 0.0067      & \textbf{-} & \textbf{-}   \\
		1552A  & 800                  & 0.211                                      & 0.0044      & 0.2116                               & 0.004       & \textbf{-} & \textbf{-}   \\
		1552B  & 1500                 & 0.4594                                     & 0.0133      & {\cellcolor[rgb]{1,0.8,0.404}}0.3199 & 0.0095      & 0.1395     & 11.99*   \\
		1553A  & 800                  & 0.2116                                     & 0.0067      & 0.2092                               & 0.0111      & \textbf{-} & \textbf{-}   \\
		1553H  & 2900                 & {\cellcolor[rgb]{0.753,0.753,0.753}}5.8183 & 0.0063      & \textbf{-}                           & \textbf{-}  & \textbf{-} & \textbf{-}   \\
		1554A  & 800                  & 0.3241                                     & 0.0064      & 0.327                                & 0.0089      & \textbf{-} & \textbf{-}   \\
		1554B  & 1700                 & 0.2195                                     & 0.002       & \textbf{-}                           & \textbf{-}  & \textbf{-} & \textbf{-}   \\
		1554C  & 1800                 & {\cellcolor[rgb]{0.753,0.753,0.753}}5.8167 & 0.0202      & {\cellcolor[rgb]{1,0.8,0.404}}0.2937 & 0.0363      & 5.523      & 186.91*  \\
		1556A  & 800                  & {\cellcolor[rgb]{0.408,0.796,0.816}}0.2158 & 0.008       & 0.2297                               & 0.0084      & 0.0139     & 1.7*     \\
		1557A  & 800                  & {\cellcolor[rgb]{0.753,0.753,0.753}}5.8175 & 0.0047      & {\cellcolor[rgb]{1,0.8,0.404}}0.2709 & 0.0041      & 5.5466     & 1250.18* \\
		1559A  & 900                  & 0.3236                                     & 0.0113      & {\cellcolor[rgb]{1,0.8,0.404}}0.2099 & 0.0053      & 0.1137     & 12.83*   \\
		1560A  & 800                  & 0.2443                                     & 0.0455      & 0.2472                               & 0.0455      & \textbf{-} & \textbf{-}   \\
		1561A  & 800                  & {\cellcolor[rgb]{0.408,0.796,0.816}}0.3008 & 0.0151      & 0.3226                               & 0.0172      & 0.0218     & 1.34*    \\
		1566A  & 800                  & 0.2296                                     & 0.0078      & 0.2292                               & 0.0075      & \textbf{-} & \textbf{-}   \\
		1566D1 & 1100                 & 0.3863                                     & 0.0142      & 0.3776                               & 0.0201      & \textbf{-} & \textbf{-}   \\
		1567E  & 2200                 & {\cellcolor[rgb]{0.753,0.753,0.753}}5.8178 & 0.0078      & \textbf{-}                           & \textbf{-}  & \textbf{-} & \textbf{-}   \\
		1569A  & 800                  & 0.2228                                     & 0.0044      & {\cellcolor[rgb]{1,0.8,0.404}}0.214  & 0.007       & 0.0088     & 1.51*    \\
		1573C  & 1800                 & {\cellcolor[rgb]{0.408,0.796,0.816}}0.4602 & 0.0041      & 0.4695                               & 0.0063      & 0.0093     & 1.74*    \\
		1574A  & 800                  & 0.2041                                     & 0.0063      & 0.205                                & 0.0101      & \textbf{-} & \textbf{-}   \\
		\bottomrule
	\end{tabular}
	}
\end{table}

\subsection{\RqTwoSen}
    \textbf{Comparison of execution time.}
    Tables \ref{tab:RQ2_time_cpp} and \ref{tab:RQ2_time_python} show the summary statistics of the execution time when receiving our generated input for generated codes and human codes written in C++ and Python, respectively.
    From our comparison, we find that 11 out of 22 C++ human codes (50\%) and 6 out of 22 (27.27\%) for Python significantly outperform generated codes (i.e., less execution times and highlighted in yellow).
    We also find that 8 out of 22 C++ generated codes (36.36\%) and 5 (22.73\%) for Python cannot execute within 5 seconds (i.e., time limit exceeded and highlighted in gray).
    On the other hand, only 3 out of 22 generated codes (13.64\%) for both C++ and Python beat human codes with less than 0.01 second difference (i.e., highlighted in green).
    From our manual investigation for the time limit exceeded cases, we find that AlphaCode introduced unnecessary nested loops in the generated codes that are from high-difficulty problems.

\begin{table}
	\centering
	\caption{\small Summary statistics and comparison of memory usages (in MB) between generated codes and human codes written in C++. Note that we show $|\Delta|$ and Cohen's d if generated codes perform significant difference (p-value $<$ 0.001).}
	\label{tab:RQ2_memory_cpp}
	\scalebox{0.79}{
	\begin{tabular}{cccccccc} 
		\toprule
		\multirow{2}{*}{} & \multirow{2}{*}{\textbf{Difficulty}} & \multicolumn{2}{c}{\textbf{Alpha Code}}                     & \multicolumn{2}{c}{\textbf{Human Code}}             & \multirow{2}{*}{$|\Delta|$} & \multirow{2}{*}{\textbf{Cohen's d}}  \\ 
		\cline{3-6}
		       &      & \textbf{Mean}                                 & \textbf{SD} & \textbf{Mean}                         & \textbf{SD} &            &              \\ 
		\midrule
		1549A  & 800  & 1.5249                                        & 0.2347      & 1.5765                                & 0.397       & \textbf{-} & \textbf{-}   \\
		1549C  & 1400 & {\cellcolor[rgb]{0.753,0.753,0.753}}14.0974   & 0.065       & 14.1612                               & 0.0914      & \textbf{-} & \textbf{-}   \\
		1551A  & 800  & {\cellcolor[rgb]{0.753,0.753,0.753}}1.5235    & 0.0386      & 1.5093                                & 0.1562      & \textbf{-} & \textbf{-}   \\
		1551B1 & 800  & 1.5273                                        & 0.2525      & 1.5041                                & 0.1566      & \textbf{-} & \textbf{-}   \\
		1552A  & 800  & 1.5762                                        & 0.3989      & 1.5416                                & 0.3124      & \textbf{-} & \textbf{-}   \\
		1552B  & 1500 & 4.9586                                        & 0.1488      & {\cellcolor[rgb]{1,0.8,0.404}}4.2563  & 0.0446      & 0.7023     & 6.36*    \\
		1553A  & 800  & 1.5039                                        & 0.1555      & 1.5303                                & 0.2427      & \textbf{-} & \textbf{-}   \\
		1553H  & 2900 & {\cellcolor[rgb]{0.753,0.753,0.753}}5.0258    & 0.055       & 15.0579                               & 0.0506      & \textbf{-} & \textbf{-}   \\
		1554A  & 800  & {\cellcolor[rgb]{0.753,0.753,0.753}}3.4807    & 0.0491      & 3.7433                                & 0.0508      & \textbf{-} & \textbf{-}   \\
		1554B  & 1700 & {\cellcolor[rgb]{0.753,0.753,0.753}}3.5278    & 0.075       & 3.5279                                & 0.0799      & \textbf{-} & \textbf{-}   \\
		1554C  & 1800 & {\cellcolor[rgb]{0.753,0.753,0.753}}\scalebox{0.85}{8194.9658} & 0.1861      & {\cellcolor[rgb]{1,0.8,0.404}}1.5427  & 0.1841      & \scalebox{.8}{8193.4231}  & \scalebox{.8}{44044.6*} \\
		1556A  & 800  & 1.5289                                        & 0.241       & 1.5258                                & 0.2391      & \textbf{-} & \textbf{-}   \\
		1557A  & 800  & 3.5021                                        & 0.0568      & 3.4982                                & 0.0478      & \textbf{-} & \textbf{-}   \\
		1559A  & 900  & 1.5283                                        & 0.257       & 1.538                                 & 0.2529      & \textbf{-} & \textbf{-}   \\
		1560A  & 800  & 1.5284                                        & 0.2379      & 1.5052                                & 0.1563      & \textbf{-} & \textbf{-}   \\
		1561A  & 800  & 1.5012                                        & 0.1567      & 1.5014                                & 0.1568      & \textbf{-} & \textbf{-}   \\
		1566A  & 800  & {\cellcolor[rgb]{0.753,0.753,0.753}}1.5178    & 0.0381      & 1.5242                                & 0.2357      & \textbf{-} & \textbf{-}   \\
		1566D1 & 1100 & 1.5177                                        & 0.2495      & 1.5085                                & 0.1564      & \textbf{-} & \textbf{-}   \\
		1567E  & 2200 & {\cellcolor[rgb]{0.753,0.753,0.753}}3.7976    & 0.0674      & 4.5731                                & 0.0806      & \textbf{-} & \textbf{-}   \\
		1569A  & 800  & 1.5225                                        & 0.2495      & 1.507                                 & 0.1554      & \textbf{-} & \textbf{-}   \\
		1573C  & 1800 & 13.635                                        & 1.3726      & {\cellcolor[rgb]{1,0.8,0.404}}12.1131 & 1.2187      & 1.5219     & 1.17*    \\
		1574A  & 800  & 1.5523                                        & 0.3235      & {\cellcolor[rgb]{1,0.8,0.404}}0.5532  & 0.062       & 0.9991     & 4.27*    \\
		\bottomrule
	\end{tabular}
	}
\end{table}

\begin{table}
	\centering
	\caption{\small Summary statistics and comparison of memory usages (in MB) between generated codes and human codes written in Python. Note that we show $|\Delta|$ and Cohen's d if generated codes perform significant difference (p-value $<$ 0.001).}
	\label{tab:RQ2_memory_python}
	\scalebox{0.8}{
	\begin{tabular}{cccccccc} 
		\toprule
		\multirow{2}{*}{} & \multirow{2}{*}{\textbf{Difficulty}} & \multicolumn{2}{c}{\textbf{AlphaCode}}                    & \multicolumn{2}{c}{\textbf{Human}}                  & \multirow{2}{*}{$|\Delta|$} & \multirow{2}{*}{\textbf{Cohen's d}}  \\ 
		\cline{3-6}
		       &      & \textbf{Mean}                               & \textbf{SD} & \textbf{Mean}                         & \textbf{SD} &            &            \\ 
		\midrule
		1549A  & 800  & {\cellcolor[rgb]{0.408,0.796,0.816}}39.201  & 0.1217      & 39.4353                               & 0.1096      & 0.2343     & 2.01*  \\
		1549C  & 1400 & {\cellcolor[rgb]{0.753,0.753,0.753}}83.2158 & 0.1259      & {\cellcolor[rgb]{1,0.8,0.404}}78.5995 & 0.1287      & 4.6163     & 36.07* \\
		1551A  & 800  & 39.1544                                     & 0.1226      & 39.188                                & 0.1253      & \textbf{-} & \textbf{-} \\
		1551B1 & 800  & 39.1969                                     & 0.1079      & 39.1793                               & 0.1105      & \textbf{-} & \textbf{-} \\
		1552A  & 800  & 39.1747                                     & 0.122       & 39.1938                               & 0.1168      & \textbf{-} & \textbf{-} \\
		1552B  & 1500 & 53.8854                                     & 0.1133      & 53.885                                & 0.1344      & \textbf{-} & \textbf{-} \\
		1553A  & 800  & 39.1811                                     & 0.1125      & 39.1672                               & 0.1248      & \textbf{-} & \textbf{-} \\
		1553H  & 2900 & {\cellcolor[rgb]{0.753,0.753,0.753}}110.542 & 0.1253      & \textbf{-}                            & \textbf{-}  & \textbf{-} & \textbf{-} \\
		1554A  & 800  & 55.1097                                     & 0.1666      & 55.1638                               & 0.1779      & \textbf{-} & \textbf{-} \\
		1554B  & 1700 & 50.6909                                     & 0.1316      & \textbf{-}                            & \textbf{-}  & \textbf{-} & \textbf{-} \\
		1554C  & 1800 & {\cellcolor[rgb]{0.753,0.753,0.753}}39.1715 & 0.1258      & 39.1646                               & 0.124       & \textbf{-} & \textbf{-} \\
		1556A  & 800  & 39.4315                                     & 0.1109      & {\cellcolor[rgb]{1,0.8,0.404}}39.1876 & 0.106       & 0.2439     & 2.24*  \\
		1557A  & 800  & {\cellcolor[rgb]{0.753,0.753,0.753}}50.1959 & 0.1177      & 58.7845                               & 0.1173      & \textbf{-} & \textbf{-} \\
		1559A  & 900  & 39.1851                                     & 0.126       & 39.1881                               & 0.1183      & \textbf{-} & \textbf{-} \\
		1560A  & 800  & 39.1883                                     & 0.1046      & 39.1852                               & 0.1104      & \textbf{-} & \textbf{-} \\
		1561A  & 800  & 39.1639                                     & 0.1255      & 39.1803                               & 0.1248      & \textbf{-} & \textbf{-} \\
		1566A  & 800  & 39.1911                                     & 0.1101      & 39.1841                               & 0.1201      & \textbf{-} & \textbf{-} \\
		1566D1 & 1100 & 41.4968                                     & 0.1182      & {\cellcolor[rgb]{1,0.8,0.404}}39.1843 & 0.1131      & 2.3125     & 19.88* \\
		1567E  & 2200 & {\cellcolor[rgb]{0.753,0.753,0.753}}62.6948 & 0.1079      & \textbf{-}                            & \textbf{-}  & \textbf{-} & \textbf{-} \\
		1569A  & 800  & 39.1977                                     & 0.1198      & 39.1744                               & 0.1205      & \textbf{-} & \textbf{-} \\
		1573C  & 1800 & 74.2036                                     & 0.1187      & {\cellcolor[rgb]{1,0.8,0.404}}65.5114 & 0.1295      & 8.6922     & 69.62* \\
		1574A  & 800  & 39.1903                                     & 0.1124      & 39.1848                               & 0.1237      & \textbf{-} & \textbf{-} \\
		\bottomrule
	\end{tabular}
	}
\end{table}

    \textbf{Comparison of memory usage.}    
    Tables \ref{tab:RQ2_memory_cpp} and \ref{tab:RQ2_memory_python} show the summary statistics of the memory usage for generated codes and human codes written in C++ and Python, respectively.
    For the limit exceeded cases (i.e., highlighted in gray), the memory usages do not represent the total memory to solve each problem, but the usages before the process got terminated.
    From our comparison, we find that 4 out of 22 C++ and Python human codes (18.18\%) significantly use a smaller amount of memory compared to generated codes (highlighted in yellow).
    Interestingly, the C++ generated code for the problem \texttt{1554C} uses over 8,000 MB.
    On the other hand, only one Python generated code significantly beats human codes with less than 1 MB difference (i.e., highlighted in green).
    Our manual investigation for the large memory usage cases shows that AlphaCode did not optimize the variable type (i.e., use \textit{long long} instead of \textit{int}) and allocated unnecessary variables (i.e., generated \textit{list} of integer, but did not use it).
    Similar to the execution time, these cases usually are high-difficulty problems.

\begin{tcolorbox}
	\textbf{Answer to \RqTwo:} No, in terms of execution time and memory usage, generated codes perform on par with or worse than human codes.
	We find that AlphaCode employs excessive nested loops and unnecessary variable declarations to solve the problems.
\end{tcolorbox}

\section{Limitations and Threats to Validity}
\label{sec:threats}
\textbf{Limitations.} 
A key limitation of this work is that we do not have access to the model of AlphaCode, which limit us to replicate the entire training and testing process for our research.
We also do not have the access to a complete list of generated solutions from their official website, which limits our ability to understand the characteristic of generated codes.
However, the competitive programming dataset for machine learning is available on GitHub \citep{Web:CodeContestsData}, which could be useful for further investigation of AlphaCode learning process.

\textbf{Threats to internal validity.} 
The first threat is the correctness of test case for evaluation.
Due to the limitations of Codeforces, we cannot retrieve the edge test cases with the maximum input size.
Instead, we generated those edge test cases based on the problem descriptions, with a manual check to minimize a threat.
The second threat is the selection of generated codes.
Due to the limited access of AlphaCode model, we can only collect the randomly selected generated codes from AlphaCode official website.
Even though those codes were randomly selected, in some cases having performance issues, the result shows that the similarity to human codes is in the same trend regardless of their performance and language.

\textbf{Threats to external validity.} 
The main external threat is the generality of results to other code generation systems.
In this study, we investigated AlphaCode, which is mainly focused on generating competitive programming code.
As a result, our findings may not apply to other types of code generation systems, such as GitHub Copilot.
However, our approaches can be applied to those systems for further evaluation.
Another threat is the sample size of the analyzed data.
We analyzed only 44 generated codes with 31,736 human codes due to the dataset limitations.
This small sample might not be able to represent the population.
However, we manually checked those data to ensure the quality and to reduce bias.

\section{Conclusions and Future Works}
\label{sec:conclusion}
This study conducted an empirical analysis to examine the performance and comparability of the AlphaCode-generated codes to human codes.
Our results show that (i) AlphaCode generates similar codes to humans, which are comprised of various human code fragments and
(ii) the generated code performs on par with or worse than the human code in terms of execution time and memory utilization.
These results indicate that software developers should review the generated codes as they might contain problematic codes and introduce performance issues.

In future work, we want to extend the study to a larger dataset including more languages and problems.
While we have used source code similarity, source code naturalness~\cite{rahman_natural_2019} might be an interesting metric to understand the generated codes.
We are also interested in other code generation models such as GitHub Copilot~\cite{Web:GitHubCopilot}.

\section*{Acknowledgment}
This work has been supported by Japan Society for the Promotion of Science (JSPS) KAKENHI Grant Numbers JP20H05706 and JP20K19774.

\bibliographystyle{IEEEtranN}
\bibliography{bibliography.bib}

\newcolumntype{C}{>{\centering\arraybackslash}X} 
\setlength{\extrarowheight}{1pt}

\end{document}